# Genetic Analysis about Differentiation of Helper T Lymphocytes

Qixin Wang[1*], Menghui Li[2], Li Charlie Xia[3], Ge Wen[4], Hualong Zu[5], Mingyi Gao[6]

[1*]*Department of Mathematics, University of Southern California, Los Angeles, CA 90089, USA*

[2]*Department of Biomedical Engineering, Peking University, Beijing, 100871, PR China*

[3]*Department of Biological Sciences, University of Southern California, Los Angeles, CA 90089, USA*

[4]*Department of Epidemiology, University of Southern California, Los Angeles, CA 90089, USA*

[5]*Department of Electrical Engineering, University of Southern California, Los Angeles, CA 90089, USA*

[6]*Department of Chemical Engineering and Materials, University of Southern California, Los Angeles, CA 90089, USA*

[1*]*Corresponding author: qixin.wang@usc.edu*

**ABSTRACT.** In human immune system, helper T lymphocytes are able to differentiate into two lymphocyte subsets: Th1 and Th2. The intracellular signaling pathways of differentiation form a dynamical regulation network by means of secreting distinctive types of cytokines, while the differentiation is regulated by two major gene loci: T-bet and GATA-3. In this article, a system dynamics model is conducted to simulate the differentiation and re-differentiation process of helper T lymphocytes, based on the experimental data of the gene expression levels of T-bet and GATA-3 during the differentiation process of helper T lymphocytes. The follow-up discussion covers three ultimate states of the model and then arrives at the conclusion that cell differentiation potential exists while system dynamic model at the unstable equilibrium point; the helper T lymphocyte will no longer have the potential of differentiation when the model evolution reached a stable equilibrium point. In addition, the time lag, caused by the expression of transcription factors, can lead to the oscillation phenomenon of the secretion of cytokines during differentiation phase.

**1. Introduction**

CD4 T helper cells, playing a crucial role in immune reaction, are ubiquitously known as a significant component of the human immune system. Human T helper cells are further categorized into two subpopulations: Th1 cells and Th2 cells [1]. The signaling transduction network that regulated Th1-Th2 differentiation is a typical dynamical network system.

In their process of differentiation, Th1 cells secrete cytokines such as $TNF-\beta$, IL-12 and IL-18, as well as $IFN-\gamma$ [2-5], while Th2 cells produce IL-4 (massively), IL-5, IL-6, IL-9 and IL-10[6,7]. It is through secreting various cytokines that they function correctly as helpers in immune responses of distinct types.

With continual breakthroughs of researches on T cells, the mechanism of Th1 and Th2 differentiation unfolds gradually. $CD4^+$ cells, when remaining undifferentiated, only secrete a tiny amount of cytokines and they are labeled as Th0 cells or naive T helper precursor cells [8,9,10,11]. From the two differentiation paths, they will finally choose one, and their choice is decided on the external environment (various cytokines), mainly by $IFN-\gamma$ and IL-4 [12, 13, 14].

External $IFN-\gamma$ can induce Th0 to differentiate into Th1. Firstly, $IFN-\gamma$ binds to its receptors on the outer surface of cell membrane, which activates the signal molecule STAT-1 and triggers nuclear translocation of it [15, 16]. After that, STAT-1 activates the transcription factor T-bet, which plays a dominant role in regulating differentiation of helper T cells [17]. When T-bet binds to the gene that can be transcribed and translated into $IFN-\gamma$, histone acetylation and DNA methylation are catalyzed and the gene is modified, hence the alteration of its accessibility, which leads to chromatin remodeling [18-21]. Generally, the $IFN-\gamma$ gene undergoes morphological changes so that it can be transcribed more easily. As the production of $IFN-\gamma$ is augmented, part of it is transported out of the cell and once again activates the $IFN-\gamma$/T-bet pathway to establish a positive feeding looping [22-25]. Moreover, T-bet can also influence GATA-3's function to inhibiting the transcription of IL-4[16].

Besides, IL-12, mainly secreted by dendritic cells, monocytes and macrophages, is another crucial cytokine that induces Th1 differentiation. At early differentiation stage, IL-12 activates STAT4 via the receptor complex on the cell membrane [5]. STAT4 can simultaneously affect the $IFN-\gamma$ gene and the T-bet gene [18, 19], leading to more combination of $IFN-\gamma$ and T-bet, which boosts the translation of $IFN-\gamma$ and its production. A certain period of time after the onset of differentiation, differentiation can proceed smoothly even without IL-12, because the necessity of its promoting effect decreases as time elapses [8,20]. Note that the IL-12 is not an essential cytokine: when the IL-12 gene of T help cells is knocked out, they can still secretes $IFN-\gamma$, but of much lower level than normal [26].

The external IL-4 is able to induce Th0 to differentiate into Th2 [7, 27, 28]. Initially, IL-4 binds to its receptors on the outer surface of cell membrane, and activates STAT-6 and finally GATA-3[9,27-30 ]. Then, the activated GATA-3 changes the accessibility of IL-4 and therefore stimulates the transcription of IL-4, forming a positive feedback looping. GATA-3 can also inhibit the function of T-bet in the same way as T-bet does[28,29,30].

The process of differentiation can be divided into two stages: polarization and differentiation. Polarization is defined as the structural transformation of Th0 towards either subpopulation of Th1 and Th2 prior to the completion of differentiation. The structural change of Th0 at this stage is reversible, which means the cell is still able to switch from Th1 to Th2 or backwards, if its culture environment is properly altered. However, a Th2 cell with thorough differentiation cannot possibly transform into a Th1 cell, even when a large quantity of $IFN-\gamma$ is added to the culture environment or a high level of T-bet is expressed in cells. Nevertheless, the latter method does generate a special type of cell that produces both IL-4 and $IFN-\gamma$, which is presumably due to the subtle change that the IL-4 gene with chromatin remodeling is no longer inhibited by T-bet[30,31,32].

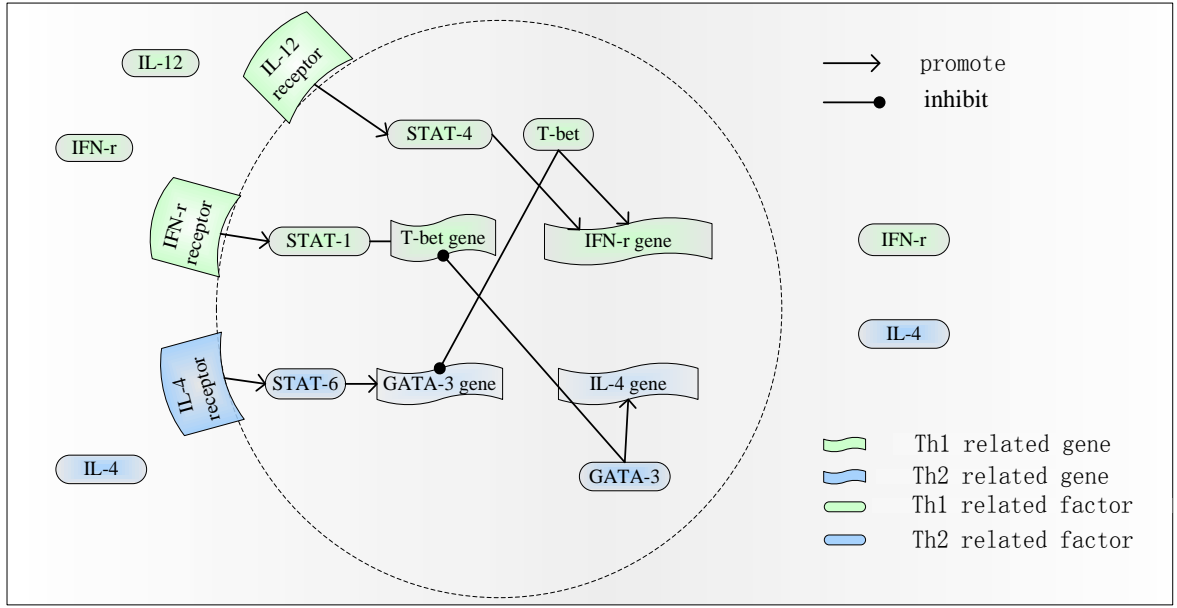

Figure 1: The differentiation-related intracellular signaling pathway for Th1 and Th2 cells.

## 2. Materials and Methods

### 2.1 Model Building

System dynamic models have a wide range of applications in the field of biological regulatory networks [34-36]. We developed a lag-system dynamical model [33] to simulate the differentiation process of Th0 cell to Th1/Th2 cells:

2.1.1 The differentiation of Th1:

Firstly, if we define the concentration of $T-bet$ as $T(i)$ at the time $i$, the production rate of $IFN-\gamma$ (i.e., $v_{IFN-\gamma(i)produce}$) can be modeled as

$$v_{IFN-\gamma(i)produce} = \frac{k_{v1}T(i)}{k_{1m} + T(i)} \tag{1}$$

Where $k_{v1}$ and $k_{1m}$ are meaningful coefficients here. Eq. (1) apparently shows that $v_{IFN-\gamma(i)produce}$ increases non-linearly as the concentration of $T-bet$ augments, and have $v_{IFN-\gamma(i)produce} \to \frac{k_{v1}T(i)}{k_{1m}}$ when $T(i) \to 0$, which means $v_{IFN-\gamma(i)produce}$ is almost directly proportional to $T(i)$, when the concentration of $T-bet$ is extremely low. $k_{1m}$ denotes the binding capacity of $T-bet$ and the $IFN-\gamma$ gene (the smaller $k_{1m}$ is, the stronger the binding capacity is, and vice versa)

I. We get $v_{IFN-\gamma(i)produce} \to k_{v1}$, when $T(i) \to \infty$. In the condition that the concentration of $T-bet$ is elevated to an exceptionally high level, $v_{IFN-\gamma(i)produce}$ approaches to a fixed value (i.e. $k_{v1}$). In another word, $k_{v1}$ represents the production rate of $IFN-\gamma$, when the $IFN-\gamma$ gene is being transcribed continuously. Due to the fact that the transduction pathway is comparatively long, the concentration of $T-bet$ is actually determined by the concentration of $IFN-\gamma$ time $\tau_2$ ago (i.e., the point in time $i-\tau_2$, or $\beta^{\tau_2}i$, since we have $\beta^{\tau}A(i) = A(i-\tau)$, where $\beta$ is the delay operator about $i$). Here $\tau_2$ is in fact the non-negligible period of time that $IFN-\gamma$ needs to stimulate $T-bet$. Hence the production rate of $T-bet$ is expressed by:

$$v_{T(i)produce} = \frac{k_{IF}\beta^{\tau_2}v_{IFN-\gamma(i)produce}}{1+\frac{k_{IF}}{k_{IFT}}\beta^{\tau_2}v_{IFN-\gamma(i)produce}} \tag{2}$$

Further analysis of Eq.(2) reveals the following information:

i. when $v_{IFN-\gamma(i)produce} \to 0$ and $v_{IFN-\gamma(i)produce} \ll v_{T(i)produce}$, $v_{T(i)produce}$ approaches to $k_{IF}\beta^{\tau_2}v_{IF-\gamma(i)produce}$. That is to say, given a low production rate of $IFN-\gamma$, the production rate of $T-bet$ is to a very large extent decided by $v_{IF-\gamma(i)produce}$ and $k_{IF}$, which are the production rate of $IFN-\gamma$ and the amplification factor of the signal cascade, respectively.

ii. When $v_{IFN-\gamma(i)produce} \to k_{v1}$ and $v_{IFN-\gamma(i)produce} \leq v_{T(i)produce}$, we have $v_{T(i)produce} \to \frac{k_{IFT}k_{IF}k_{v1}}{k_{IFT}+k_{IF}k_{v1}}$, which means that the production rate of $T-bet$ approaches to a fixed value $\dfrac{k_{IFT} k_{IF} k_{v1}}{k_{IFT} + k_{IF} k_{v1}}$, when $v_{IFN-\gamma(i)produce}$ is close to $k_{v1}$ and at the same time not far from $v_{G(i)produce}$.

iii. When $v_{IFN-\gamma(i)produce} \to k_{v1}$ and $v_{IFN-\gamma(i)produce} \gg v_{T(i)produce}$, we will expect to have $v_{T(i)produce} \to k_{IFT}$. When the production rate of $IFN-\gamma$ is much higher than that of $T-bet$, the $T-bet$ production rate will approach to, but is always lower than $k_{IFT}$, due to the speed limitation of $T-bet$ transcription.

If we substitute Eq.(1) into Eq.(2), we will arrive at:

$$v_{T(i)produce} = \dfrac{k_{IF} \dfrac{k_{v1} \beta^{\tau_2} T(i)}{k_{1m} + \beta^{\tau_2} T(i)}}{1 + \dfrac{k_{IF}}{k_{IFT}} * \dfrac{k_{v1} \beta^{\tau_2} T(i)}{k_{1m} + \beta^{\tau_2} T(i)}} \quad (3)$$

Its simplified form:

$$v_{T(i)produce} = \dfrac{k_{IF} k_{v1} k_{IFT} \beta^{\tau_2} T(i)}{k_{1m} k_{IFT} + (k_{IFT} + k_{IF} k_{v1}) \beta^{\tau_2} T(i)} \quad (4)$$

Since $GATA-3$ can affect the binding ability of $T-bet$ to the $IFN-\gamma$ gene, the binding rate of $T-bet$ and $IFN-\gamma$ is given by the following formulas:

$$v_{T(i)bind} = \dfrac{k_{vT} T(i)}{k_{mT} + T(i)} \quad (5)$$

$$k_{mT} = k_{Tm}(1 + \dfrac{G(i) I_B}{k_{Gm}})(1 + \dfrac{k_c C_{IL-12}}{i}) \quad , \quad I_B = \begin{cases} 1 & , i \notin B \\ 0 & , i \in B \end{cases} \quad B = \left\{ i \middle| T(i) = \dfrac{1}{n_B} \sum_{\tau=i-n_B}^{i-1} \beta^{\tau} T(i) \right\} \quad (6)$$

where $k_{mT}$ represents the binding ability of $T-bet$ to genes with the presence of $GATA-3$; $k_{Gm}$ represents the binding ability of $GATA-3$ to genes without the presence of $T-bet$; $k_{Tm}$ represents the binding ability of $T-bet$ to genes without the presence of $GATA-3$; $C_{IL-12}$ is the concentration of $IL-12$ at early differentiation stage.

As $G(i)$ augments, it inhibits $T-bet$ more significantly, and when less $GATA-3$ produce, more $T-bet$ will binds to the $IFN-\gamma$ gene, hence $k_{Gm}$ and $k_{mT}$ are negatively correlated. If $T(i)$ remains approximately the same value for a rather long period of time (i.e., $n_B$), it means that the Th cells have reached the stage of differentiation completion. Under such circumstances the process of differentiation is irreversible even when a large quantity of $IL-4$ is added to the culture environment or a high level of $GATA-3$ is expressed in cells, in another word, the change of $G(i)$ cannot affect the binding ability of $T-bet$ to the $IFN-\gamma$ gene. At the early stage of Th1 differentiation, $IL-12$ is able to simultaneously stimulate both $IFN-\gamma$ and T-bet gene through $IL-12 \to STAT-4$ transduction pathway and accelerates the $IFN-\gamma$ gene transcription. As the process of differentiation proceeds, the promoting effect of $IL-12$ diminishes, till differentiation progresses self-reliantly even without the presence of $IL-12$ (when $i \to \infty, 1+\dfrac{k_c C_{IL-12}}{i} \to 1$ ).

We substitute Eq. (6) into Eq. (5) to get:

$$v_{T(i)bind} = \dfrac{k_{vT} T(i)}{k_{Tm}(1+\dfrac{G(i)I_B}{k_{Gm}})(1+\dfrac{k_c C_{IL-12}}{i}) + T(i)} \tag{7}$$

Then we assume that $T-bet$ won't degrade before binding to the $IFN-\gamma$ gene, based on which we have:

$$\dfrac{dT(i)}{dt} = v_{T(i)produce} - v_{T(i)bind} \tag{8}$$

We substitute Eq. (4), Eq.(7) into Eq.(8), so we have:

$$\dfrac{dT(i)}{dt} = \dfrac{k_{IF} k_{v1} k_{IFT} \beta^{\tau_2} T(i)}{k_{1m} k_{IFT} + (k_{IFT} + k_{IF} k_{v1}) \beta^{\tau_2} T(i)} - \dfrac{k_{vT} T(i)}{k_{Tm}(1+\dfrac{G(i)I_B}{k_{Gm}})(1+\dfrac{k_c C_{IL-12}}{i}) + T(i)} \tag{9}$$

## 2.1.2 The differentiation of Th2:

We define $G(i)$ as the concentration of $GATA-3$ in cells at the moment $i$, and the production rate of $IL-4$ is modeled as:

$$v_{IL-4(i)produce} = \frac{k_{v6}G(i)}{k_{6m} + G(i)} \tag{10}$$

where $k_{v1}$ and $k_{1m}$ are meaningful coefficients here. Eq. (10) shows that $v_{IL-4(i)produce}$ augments non-linearly with the increase of concentration of $GATA-3$, and:

I. When $G(i) \to 0$, we have $v_{IL-4(i)produce} \to \frac{k_{v6}G(i)}{k_{6m}}$, which means $v_{IL-4(i)produce}$ is almost directly proportional to $G(i)$, when the concentration of $GATA-3$ is extremely low. $k_{6m}$ denotes the binding capacity of $GATA-3$ to the $IL-4$ gene (the smaller $k_{6m}$ is, the stronger the binding capacity is, and vice versa)

II. When $G(i) \to \infty$, we have $v_{IL-4(i)produce} \to k_{v6}$. In the condition that the concentration of $GATA-3$ is elevated to an exceptionally high level, $v_{IL-4(i)produce}$ approaches to a fixed value (i.e. $k_{v6}$). In another word, $k_{v6}$ represents the maximum production rate of $IL-4$, when the $IL-4$ gene is being transcribed continuously. Due to the fact that the transduction pathway is comparatively long, the concentration of $GATA-3$ is actually determined by the concentration of $IL-4$ time $\tau_1$ ago (i.e., the point in time $i-\tau_1$, or). Here $\tau_1$ is in the non-negligible time delay that $IL-4$ needs to stimulate $GATA-3$. Hence the production rate of $GATA-3$ is expressed by:

$$v_{G(i)produce} = \frac{k_{IL}\beta^{\tau_1}v_{IL-4(i)produce}}{1 + \frac{k_{IL}}{k_{ILG}}\beta^{\tau_1}v_{IL-4(i)produce}} \tag{11}$$

Further analysis of Eq.(11) reveals the following information:

i. when $v_{IL-4(i)produce} \to 0$ and $v_{IL-4(i)produce} \ll v_{G(i)produce}$, $v_{G(i)produce}$ approaches to $k_{IL}\beta^{\tau_1}v_{IL-4(i)produce}$, which means, given a low production rate of $IL-4$, the production rate of $GATA-3$ is to a very large extent decided by $v_{IL-4(i)produce}$ and $k_{IL}$, which are the production rate of $IL-4$ and the amplification factor of the signal cascade, respectively.

ii. When $v_{IL-4(i)produce} \to k_{v6}$ and $v_{IL-4(i)produce} \leq v_{G(i)produce}$, we have $v_{G(i)produce} \to \frac{k_{ILG}k_{IL}k_{v6}}{k_{ILG}+k_{IL}k_{v6}}$, which means that the production rate of $GATA-3$ approaches to a fixed value $\frac{k_{ILG}k_{IL}k_{v6}}{k_{ILG}+k_{IL}k_{v6}}$, when $v_{IL-4(i)produce}$ is close to $k_{v6}$ and at the same time not far from $v_{T(i)produce}$.

iii. When $v_{IL-4(i)produce} \to k_{v6}$ and $v_{IL-4(i)produce} \gg v_{G(i)produce}$, we will expect to have $v_{G(i)produce} \to k_{ILG}$. When the production rate of $IL-4$ is much higher than that of $GATA-3$, the $GATA-3$ production rate will approach to, but is always lower than $k_{ILG}$, due to the speed limitation of $T-bet$ transcription.

We substitute Eq.(10) into Eq.(11), then arrive at:

$$v_{G(i)produce} = \frac{k_{IL}\frac{k_{v6}\beta^{\tau_1}G(i)}{k_{6m}+\beta^{\tau_1}G(i)}}{1+\frac{k_{IL}}{k_{ILG}}*\frac{k_{v6}\beta^{\tau_1}G(i)}{k_{6m}+\beta^{\tau_1}G(i)}} \qquad (12)$$

And its simplified form is:

$$v_{G(i)produce} = \frac{k_{IL}k_{v6}k_{ILG}\beta^{\tau_1}G(i)}{k_{6m}k_{ILG}+(k_{ILG}+k_{IL}k_{v6})\beta^{\tau_1}G(i)} \qquad (13)$$

Since $T-bet$ can affect the binding ability of $GATA-3$ to the $IL-4$ gene, the binding rate of $GATA-3$ and $IL-4$ is given by the following formulas:

$$v_{G(i)bind} = \frac{k_{vG}G(i)}{k_{mG}+G(i)} \qquad (14)$$

$$k_{mG} = k_{Gm}(1+\frac{T(i)I_A}{k_{Tm}}), \quad I_A = \begin{cases} 1 & ,i \notin A \\ 0 & ,i \in A \end{cases} \quad A = \left\{ i \middle| G(i) = \frac{1}{n_A} \sum_{\tau=i-n_A}^{i-1} \beta^{\tau} G(i) \right\} \quad (15)$$

Where $k_{mG}$ represents the binding ability of $GATA-3$ to $IL-4$ gene with the presence of $T-bet$; $k_{Gm}$ represents the binding ability of $GATA-3$ to $IL-4$ gene without the presence of $T-bet$; $k_{Tm}$ represents the binding ability of $T-bet$ to $IL-4$ gene without the presence of $GATA-3$.

As $T(i)$ augments, it inhibits $GATA-3$ more significantly, and when less $T-bet$ produce, more $GATA-3$ will binds to the $IL-4$ gene, hence $k_{Tm}$ and $k_{mG}$ are negatively correlated. If $G(i)$ remains approximately the same value for a rather long period of time (i.e. $n_A$), it shows that the Th cells have reached the stage of differentiation completion. Under such circumstances the process of differentiation is irreversible even when a large quantity of $IFN-\gamma$ is added to the culture environment or a high level of $T-bet$ is expressed in cells, in another word, the change of $T(i)$ cannot affect the binding ability of $GATA-3$ to the $IL-4$ gene any more.

We substitute Eq.(15) into Eq.(14) to get:

$$v_{G(i)bind} = \frac{k_{vG}G(i)}{k_{Gm}(1+\frac{T(i)I_A}{k_{Tm}})+G(i)} \quad (16)$$

Then we assume that $GATA-3$ won't degrade before binding to the $IL-4$ gene, based on which we have:

$$\frac{dG(i)}{dt} = v_{G(i)produce} - v_{G(i)bind} \quad (17)$$

We substitute Eq.(13), Eq.(16) into Eq.(17), so we have:

$$\frac{dG(i)}{dt} = \frac{k_{IL}k_{v6}k_{ILG}\beta^{\tau_1}G(i)}{k_{6m}k_{ILG}+(k_{ILG}+k_{IL}k_{v6})\beta^{\tau_1}G(i)} - \frac{k_{vG}G(i)}{k_{Gm}(1+\frac{T(i)I_A}{k_{Tm}})+G(i)} \quad (18)$$

Note that Th cells is almost unable to spontaneously secrete cytokines except $IFN-\gamma$ and $IL-4$ without the stimulation of other types of cytokines. Since this experiment is designed to culture Th cells in the environment where no cytokines exist but, $IFN-\gamma$, $IL-12$ and $IL-4$, the possible existence and influence of other cytokines can be theoretically ignored. Thus, the differentiation model of Th cells is modeled as:

$$\begin{cases} \dfrac{dT(i)}{dt} = \dfrac{k_{IF}k_{v1}k_{IFT}T(i-\tau_2)}{k_{1m}k_{IFT}+(k_{IFT}+k_{IF}k_{v1})T(i-\tau_2)} - \dfrac{k_{vT}T(i)}{k_{Tm}(1+G(i)I_B/k_{Gm})(1+k_cC_{IL-12}/i)+T(i)} \\ \dfrac{dG(i)}{dt} = \dfrac{k_{IL}k_{v6}k_{ILG}G(i-\tau_1)}{k_{6m}k_{ILG}+(k_{ILG}+k_{IL}k_{v6})G(i-\tau_1)} - \dfrac{k_{vG}G(i)}{k_{Gm}(1+T(i)I_A/k_{Tm})+G(i)} \end{cases} \quad (19)$$

## 2.2 Model Analysis

The completion of the differentiation process is reached when:

$$\begin{cases} \dfrac{k_{IF}k_{v1}k_{IFT}T(i-\tau_2)}{k_{1m}k_{IFT}+(k_{IFT}+k_{IF}k_{v1})T(i-\tau_2)} - \dfrac{k_{vT}T(i)}{k_{Tm}(1+G(i)I_B/k_{Gm})+T(i)} = 0 \\ \dfrac{k_{IL}k_{v6}k_{ILG}G(i-\tau_1)}{k_{6m}k_{ILG}+(k_{ILG}+k_{IL}k_{v6})G(i-\tau_1)} - \dfrac{k_{vG}G(i)}{k_{Gm}(1+T(i)I_A/k_{Tm})+G(i)} = 0 \end{cases} \quad (20)$$

At this moment, limit points for the equations exist:

I. When the cells are at Th0 stage, we have $T(i)=0$ and $G(i)=0$. According to Eq.(20):

$$\begin{cases} \dfrac{k_{IF}k_{v1}k_{IFT}T(i-\tau_2)}{k_{1m}k_{IFT}+(k_{IFT}+k_{IF}k_{v1})T(i-\tau_2)} - \dfrac{k_{vT}T(i)}{k_{Tm}(1+G(i)I_B/k_{Gm})(1+k_cC_{IL-12}/i)+T(i)} = 0 \\ \dfrac{k_{IL}k_{v6}k_{ILG}G(i-\tau_1)}{k_{6m}k_{ILG}+(k_{ILG}+k_{IL}k_{v6})G(i-\tau_1)} - \dfrac{k_{vG}G(i)}{k_{Gm}(1+T(i)I_A/k_{Tm})+G(i)} = 0 \end{cases} \quad (21)$$

Thus, the ordinary differential equations (Eq.(19)) have a incipient state: $P_0(G_0^0, T_0^0)=(0,0)$

II. The differentiation of Th1 is considered to be completed, when

$i \to \infty$, $G(i) \to 0$ and $I_B = 0$. According to Eq.(19),

$$\lim_{G(i)\to 0} \frac{k_{IF}k_{v1}k_{IFT}T(i-\tau_2)}{k_{1m}k_{IFT}+(k_{IFT}+k_{IF}k_{v1})T(i-\tau_2)} - \frac{k_{vT}T(i)}{k_{Tm}(1+G(i)I_B/k_{Gm})(1+k_c C_{IL-12}/i)+T(i)}$$

$$= \frac{k_{IF}k_{v1}k_{IFT}T(i)}{k_{1m}k_{IFT}+(k_{IFT}+k_{IF}k_{v1})T(i)} - \frac{k_{vT}T(i)}{k_{Tm}+T(i)},$$

upon which we have:

$$\frac{k_{IF}k_{v1}k_{IFT}T(i)}{k_{1m}k_{IFT}+(k_{IFT}+k_{IF}k_{v1})T(i)} - \frac{k_{vT}T(i)}{k_{Tm}+T(i)} = 0 \tag{22}$$

The solution is:

$$T(i) = \frac{k_{IF}k_{v1}k_{IFT}k_{Tm} - k_{1m}k_{IFT}k_{vT}}{k_{IFT}k_{vT}+k_{IF}k_{v1}k_{vT}-k_{IF}k_{v1}k_{IFT}} \tag{23}$$

On the other hand:

$$\lim_{G(i)\to 0} \frac{k_{IL}k_{v6}k_{ILG}G(i-\tau_1)}{k_{6m}k_{ILG}+(k_{ILG}+k_{IL}k_{v6})G(i-\tau_1)} - \frac{k_{vG}G(i)}{k_{Gm}(1+T(i)I_A/k_{Tm})+G(i)} = 0 \tag{24}$$

So one limit point for Eq.(19) is：

$$P_0(G_1^0, T_1^0) = (0, \frac{k_{IF}k_{v1}k_{IFT}k_{Tm} - k_{1m}k_{IFT}k_{vT}}{k_{IFT}k_{vT}+k_{IF}k_{v1}k_{vT}-k_{IF}k_{v1}k_{IFT}})$$

III. The process of Th2 differentiation is over when $i \to \infty$, $T(i) \to 0$ and $I_A = 0$. According to Eq.(19):

$$\lim_{T(i)\to 0} \frac{k_{IL}k_{v6}k_{ILG}G(i-\tau_1)}{k_{6m}k_{ILG}+(k_{ILG}+k_{IL}k_{v6})G(i-\tau_1)} - \frac{k_{vG}G(i)}{k_{Gm}(1+T(i)I_A/k_{Tm})+G(i)}$$

$$= \frac{k_{IL}k_{v6}k_{ILG}G(i)}{k_{6m}k_{ILG}+(k_{ILG}+k_{IL}k_{v6})G(i)} - \frac{k_{vG}G(i)}{k_{Gm}+G(i)},$$

upon which we have:

$$\frac{k_{IL}k_{v6}k_{ILG}G(i)}{k_{6m}k_{ILG}+(k_{ILG}+k_{IL}k_{v6})G(i)} - \frac{k_{vG}G(i)}{k_{Gm}+G(i)} = 0 \tag{25}$$

The solution is:

$$G(i) = \frac{k_{IL}k_{v6}k_{ILG}k_{Gm} - k_{6m}k_{ILG}k_{vG}}{k_{ILG}k_{vG} + k_{IL}k_{v6}k_{vG} - k_{IL}k_{v6}k_{ILG}} \tag{26}$$

On the other hand:

$$\lim_{T(i) \to 0} \frac{k_{IF}k_{v1}k_{IFT}T(i-\tau_2)}{k_{1m}k_{IFT} + (k_{IFT} + k_{IF}k_{v1})T(i-\tau_2)} - \frac{k_{vT}T(i)}{k_{Tm}(1+G(i)I_B/k_{Gm}) + T(i)} = 0 \tag{27}$$

So another limit point for Eq. (19) is

$$P_0(G_2^0, T_2^0) = (\frac{k_{IL}k_{v6}k_{ILG}k_{Gm} - k_{6m}k_{ILG}k_{vG}}{k_{ILG}k_{vG} + k_{IL}k_{v6}k_{vG} - k_{IL}k_{v6}k_{ILG}}, 0)$$

∴ The incipient point for Eq.(19) is $P_0(G_0^0, T_0^0) = (0,0)$     (Th0 state)

The limit points for Eq.(19) are

$$P_0(G_1^0, T_1^0) = (0, \frac{k_{IF}k_{v1}k_{IFT}k_{Tm} - k_{1m}k_{IFT}k_{vT}}{k_{IFT}k_{vT} + k_{IF}k_{v1}k_{vT} - k_{IF}k_{v1}k_{IFT}}) \quad \text{(Th1 differentiation state)}$$

and

$$P_0(G_2^0, T_2^0) = (\frac{k_{IL}k_{v6}k_{ILG}k_{Gm} - k_{6m}k_{ILG}k_{vG}}{k_{ILG}k_{vG} + k_{IL}k_{v6}k_{vG} - k_{IL}k_{v6}k_{ILG}}, 0) \quad \text{(Th2 differentiation state)}$$

## 3. Results and Discussion

Mathematical modeling plays an important role in analysis of a great deal of complicated experimental phenomena of biological systems [37-44]. In this paragraph, we would like to discuss the relationship between differentiation potential and equilibrium points of the Th1-Th2 system dynamical model.

### 3.1 Time-Lag oscillation

3.1.1 Th1/Th2 differentiation process without inhibit of major control genes where $G(0) = 0, T(0) = 0 \quad \tau_1 = \tau_2 = 0$.

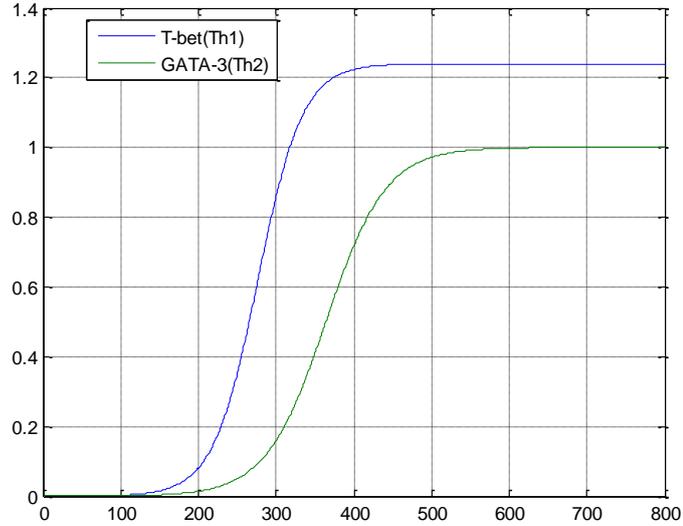

Figure 2: Expression level of major regulation gene in Th1/Th2 differentiation process without time lag. Note: Both differentiation curves on the same figure.

There is no oscillation at all and the equilibrium point is $P_0(G_1^0, T_1^0) = (0, 1.22)$ (Th1 differentiation) or $P_0(G_2^0, T_2^0) = (1, 0)$ （Th2 differentiation）.

3.1.2 Th1/Th2 differentiation process without inhibit of major control genes, where $G(0) = 0, T(0) = 0$  $\tau_1 = \tau_2 = 1$.

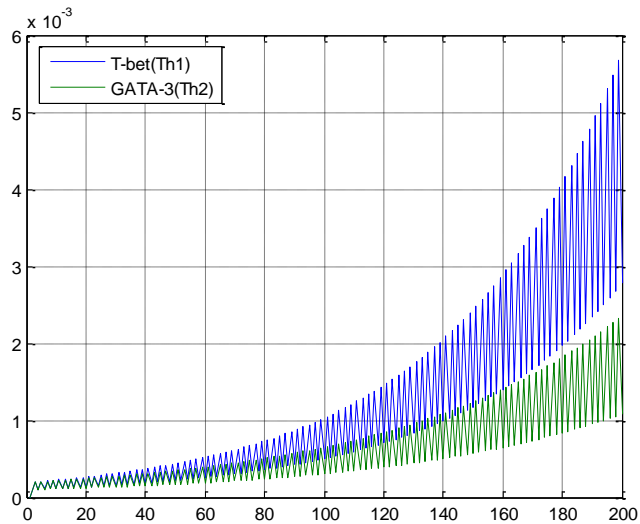

Figure 3: Expression level of major regulation gene in Th1/Th2 differentiation process with time lag in 200 iterations. Note: Both differentiation curves on the same figure.

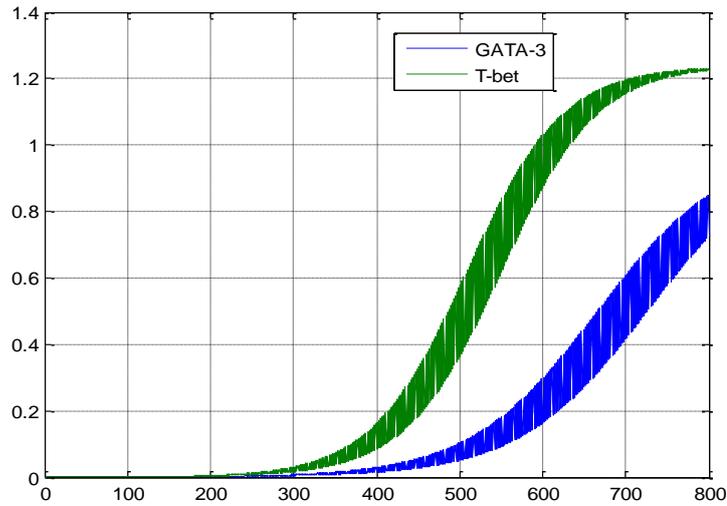

Figure 4: Expression level of major regulation gene in Th1/Th2 differentiation process with time lag in 800 iterations. Note: Both differentiation curves on the same figure.

In Figure 4, we can see that time lag oscillation occurred in time frame of 800 interactions and the equilibrium point is not reached.

3.1.3 Th1/Th2 differentiation process without inhibit of major control genes. Where $G(0)=0, T(0)=0$ $\tau_1 = \tau_2 = 5$.

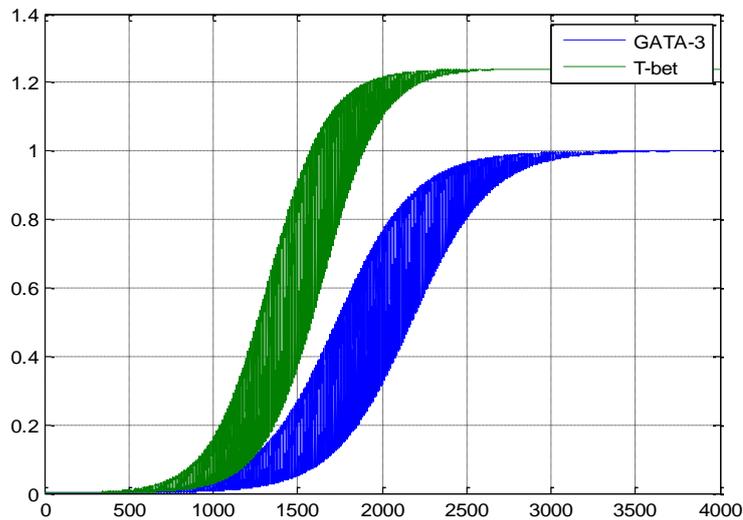

Figure 5: Expression level of major regulation gene in Th1/Th2 differentiation process with time lag in 4000 iterations. Note: Both differentiation curves on the same figure.

In Figure 5, The curve oscillates more tempestuously, and the equilibrium point is $P_0(G_1^0, T_1^0) = (0, 1.22)$ or $P_0(G_2^0, T_2^0) = (1, 0)$. Time lag oscillation occurred in time frame of 3000 iterations and the equilibrium point is reached in about 3700 iterations.

When there is no stimulation of cytokines or other signals (i.e. $G(0) = 0, T(0) = 0$), Th0 cells may differentiates into Th1 or Th2 cells with equal probability. Cytokines ($IFN-\gamma$, $IL-4$) can stimulate the transcription factors ($T-bet$, $GATA-3$) and therefore a certain period of time delay is resulted in, which brings about oscillation to the expression of $T-bet$ and $GATA-3$ (the oscillation becomes more evident if the graph is enlarged), but cannot influence the equilibrium point of the differential equation (i.e., the concentration when stable state is reached). Therefore, the oscillation can explain the abnormity of the curves representing $T-bet$, $GATA-3$ [17]. Moreover, it will take the differentiation process longer time to reach the stable point when more considerable oscillation occurs.

### 3.2 Differentiation

3.2.1 Th1 differentiation process with inhibit of major control genes.

$G(0) = 0.1, T(0) = 0.2 \quad \tau_1 = \tau_2 = 5$

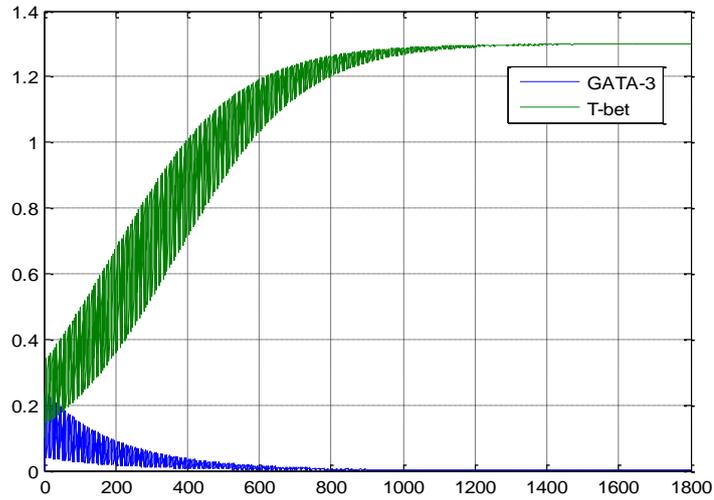

Figure 6: Expression level of major regulation gene in Th1 differentiation process with time lag. $P_0(G_1^0, T_1^0) = (0, 1.22)$

3.2.2 Th2 differentiation process with inhibit of major control genes.

$G(0) = 0.2, T(0) = 0.1 \quad \tau_1 = \tau_2 = 5$

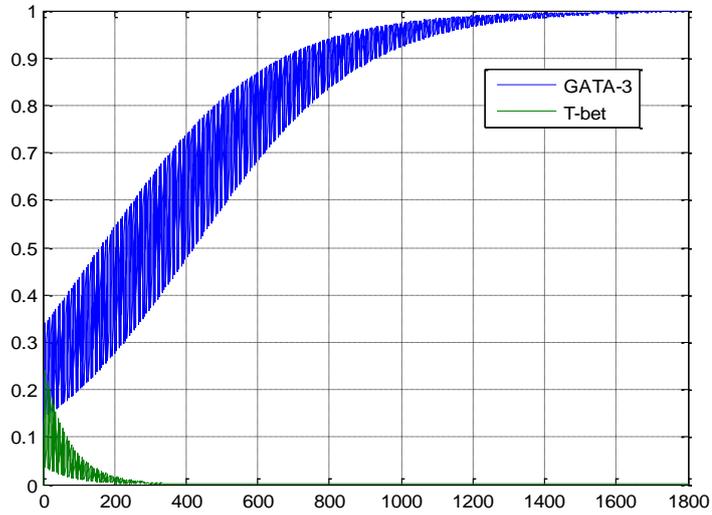

Figure 7: Expression level of major regulation gene in Th2 differentiation process with time lag. $P_0(G_2^0, T_2^0) = (1, 0)$

3.2.3 Th1/Th2 re-differentiation process with inhibit of major control genes.

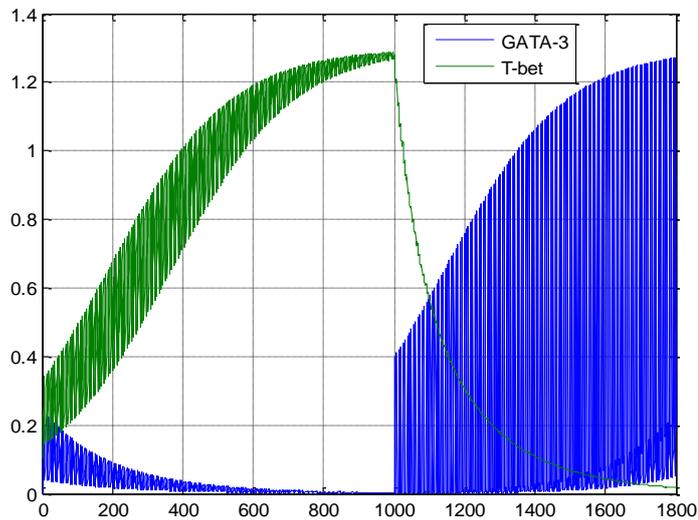

Figure 8: Expression level of major regulation gene in Th1 → Th2 re-differentiation process with time lag.

In the middle of differentiation (Figure 8), when the expression level of $T-bet$ has not reach stable equilibrium point yet, increase of expression level of $GATA-3$ will lead to the decrease of $T-bet$ expression level, which means re-differentiation occurs.

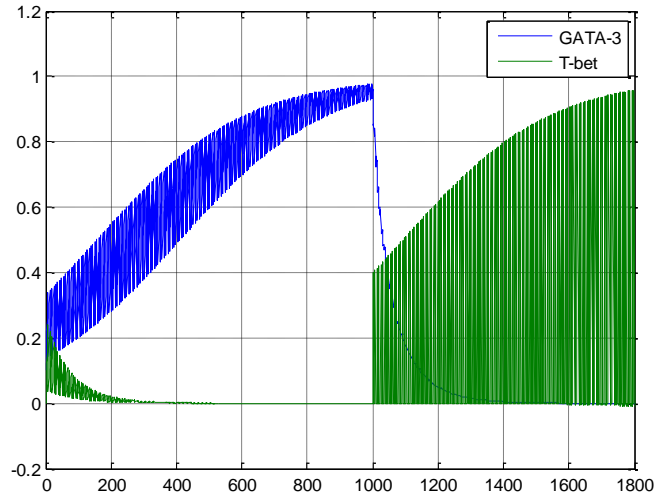

Figure 9: Expression level of major regulation gene in Th2→Th1 re-differentiation process with time lag.

In the middle of differentiation (Figure 9), when the expression level of $GATA-3$ has not reach its stable equilibrium point yet, increase of $T-bet$ expression level will lead to the decrease of $GATA-3$ expression level, which means re-differentiation occurs.

3.2.4 Th1/Th2 cell without potential of differentiation.

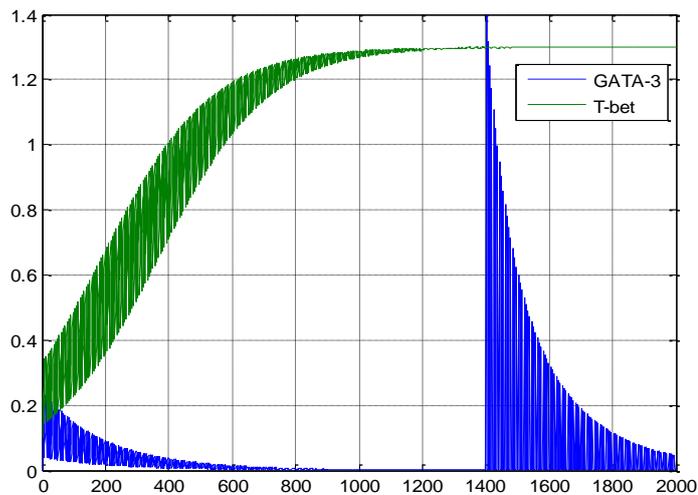

Figure 10: Th1 cell without potential of differentiation.

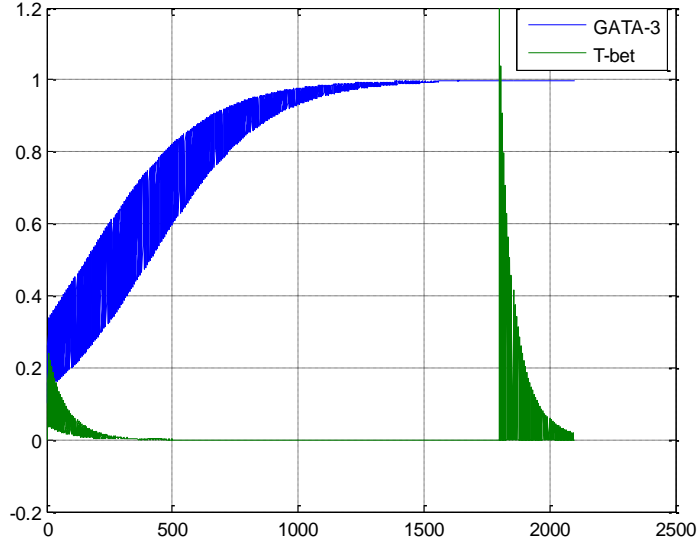

Figure 11: Th2 cell without potential of differentiation.

After through differentiation is completes (Figure 10), the expression level of $T-bet$ cannot be affected even when a high level of $GATA-3$ is expressed in cells. However, by transiently increase the expression level of $GATA-3$, it is possible to obtain re-differentiation cells that can transiently secrete $IFN-\gamma$ and $IL-4$. The mechanism is: after $G(i)$ and $T(i)$ remain stable for a while, we have $I_B \equiv 0 \Rightarrow k_{mT} \equiv k_{Tm}$ and $I_A \equiv 0 \Rightarrow k_{mG} \equiv k_{Gm}$, which helps to transform Eq.(19) into ($G(i)$ and $T(i)$ cannot affect each other):

$$\begin{cases} \dfrac{dT(i)}{dt} = \dfrac{k_{IF}k_{v1}k_{IFT}T(i-\tau_2)}{k_{1m}k_{IFT}+(k_{IFT}+k_{IF}k_{v1})T(i-\tau_2)} - \dfrac{k_{vT}T(i)}{k_{Tm}+T(i)} \\ \dfrac{dG(i)}{dt} = \dfrac{k_{IL}k_{v6}k_{ILG}G(i-\tau_1)}{k_{6m}k_{ILG}+(k_{ILG}+k_{IL}k_{v6})G(i-\tau_1)} - \dfrac{k_{vG}G(i)}{k_{Gm}+G(i)} \end{cases}$$ (Equation 30)

Due to the reconstruction of chromosome, the binding ability of $T-bet$ to the $IFN-\gamma$ gene is not influenced by $GATA-3$, and therefore the production rates of $IFN-\gamma$ and $T-bet$ remain constant. Under these circumstances, the fluctuation of $G(i)$ cannot affect $T(i)$ even when a high level of $GATA-3$ is expressed in cells.

Similarly, after through differentiation is completes (Figure 11), the expression level of $GATA-3$ cannot be affected even when a high level of $T-bet$ is expressed in cells, but it is possible to obtain re- differentiation cells that can transiently secrete $IFN-\gamma$ and $IL-4$.

3.2.5 Phase Trajectory

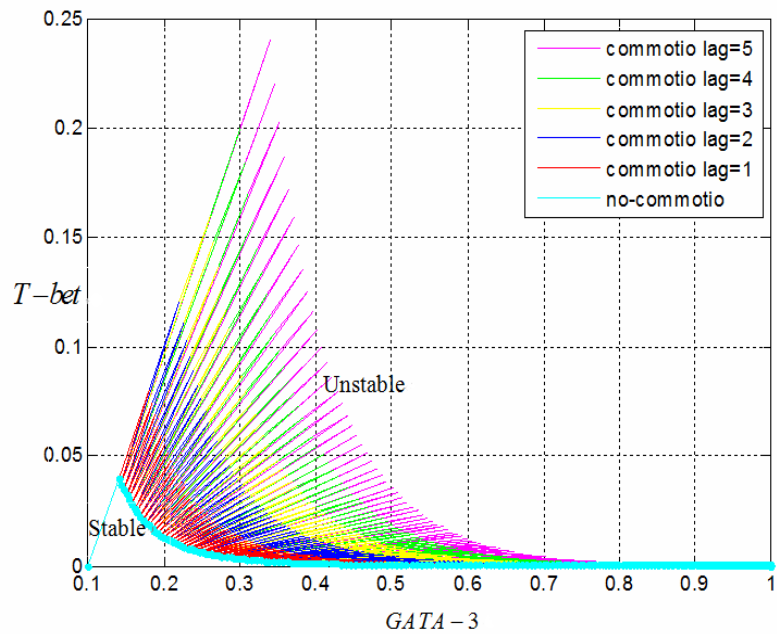

Figure 12: The phase trajectory curve of Th2 differentiation when $\tau_1 = \tau_2 = 0 \sim 5$. We can find from the figure when the $\tau$ increase, the commotion will enhance.

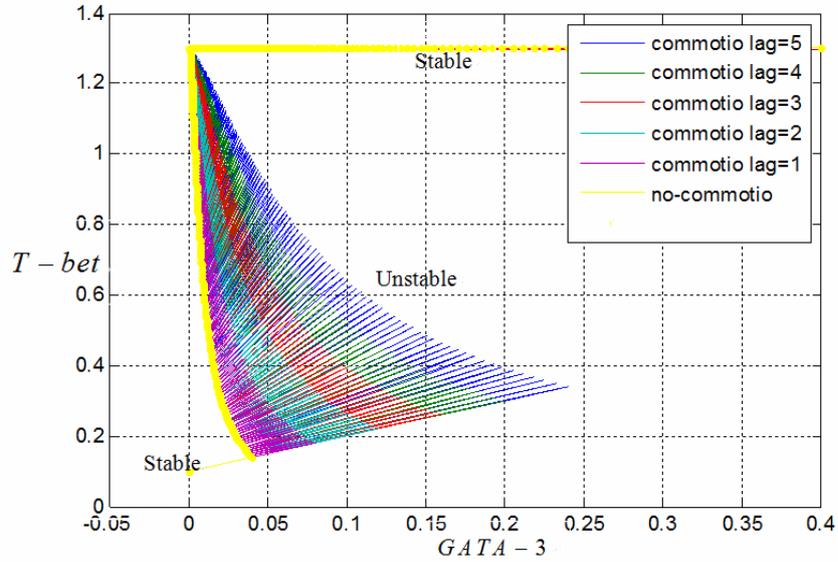

Figure 13: The phase trajectory curve of Th1 differentiation when $\tau_1 = \tau_2 = 0 \sim 5$ Note: the yellow curve in this figure is the phase trajectory of differentiation ignoring the time lag in the transduction pathway.

The Figure.12 and Figure.13 shows that oscillation doesn't exist on condition that either one of $GATA-3$ and $T-bet$ is obviously superior to the other (the comparatively stable state of differentiation), whereas in unstable state, the phase trajectory oscillates more considerably as the delay order increases. In addition, the curve oscillates violently when the level of expression of T-bet is low.

## 4. Conclusion

The helper T lymphocytes have the potential of differentiation while system dynamic model about expression of major control genes at the unstable equilibrium point. The helper T lymphocyte will no longer have the potential of differentiation when the model evolution reached a stable equilibrium point. In addition, the time lag, caused by the expression of transcription factors, can lead to the oscillation phenomenon of the secretion of cytokines during differentiation process.